\documentclass[aps,pra,twocolumn,superscriptaddress,floatfix]{revtex4}
\usepackage{graphicx}
\usepackage{dcolumn}
\usepackage{bm}
\usepackage{latexsym,epsfig}
\usepackage{graphicx}
\usepackage{verbatim}
\usepackage{comment}
\usepackage{amsmath}
\usepackage{amssymb}
\usepackage{color}
\usepackage{grffile}
\usepackage{epstopdf}

\DeclareGraphicsExtensions{.eps}

\newcommand{\beq}{\begin{equation}}
\newcommand{\eeq}{\end{equation}}
\newcommand{\bea}{\begin{eqnarray}}
\newcommand{\eea}{\end{eqnarray}}
\newcommand{\ben}{\begin{eqnarray*}}
\newcommand{\een}{\end{eqnarray*}}
\newcommand{\bfig}{\begin{figure}}
\newcommand{\efig}{\end{figure}}

\newcommand{\ra}{\rangle}
\newcommand{\la}{\langle}

\begin{document}
\title{Interacting bosons in generalized zig-zag and railroad-trestle models}

\author {Sebastian Greschner}
\affiliation{Department of Quantum Matter Physics, University of Geneva, 1211 Geneva, Switzerland} 

\author{Tapan Mishra}
\affiliation{Department of Physics, Indian Institute of Technology, Guwahati, Assam - 781039, India}

\date{\today}

\begin{abstract}
We theoretically study the ground-state phase diagram of strongly interacting bosons on a 
generalized zig-zag ladder model, the rail-road trestle (RRT) model. 
By means of analytical arguments in the limits of decoupled chains and the case of vanishing fillings 
as well as extensive DMRG calculations we examine the rich interplay between frustration and 
interaction for various parameter regimes. We distinguish three different cases, the fully 
frustrated RRT model where the dispersion relation becomes doubly degenerate and an extensive chiral 
superfluid regime is found, the anti-symmetric RRT with alternating $\pi$ and $0$ fluxes through the 
ladder plaquettes and the sawtooth limit, which is closely related to the latter case. 
We study detailed phase diagrams which include besides different single component superfluids, 
the chiral superfluid phases, the two component superfluids and different gaped phases, 
with dimer and a charge-density wave order. 
\end{abstract}

\maketitle

\section{Introduction}

Frustrated systems are one of the most interesting as well as widely explored yet still most challenging problems in the field of condensed 
matter physics. 
Frustration in one and quasi-one dimensional systems, such as quasi one-dimensional magnetic 
materials~\cite{Hase2004magnetic,Masuda2005spin,Drechsler2007frustrated, vasiliev2018milestones}, 
are of paramount importance due to the strong correlations which in interplay with the geometric frustration lead to non-trivial and intriguing physics. 
Theoretically in particular the $J_1$-$J_2$ spin model, with a frustrated next-nearest neighbour tunnelling amplitude $J_2$, has been extensively studied during the 
recent decades and important milestones include the famous analytical solution for the isotropic spin-$1/2$ $J_1$-$J_2$-model 
by Majumdar and Ghosh~\cite{Majumdar1969} or the Ising type phase transition between the critical Luttinger-liquid XY and the gapped 
dimerized~(D) phase \cite{Haldane1982, Okamoto1992}.
Detailed ground-state properties in different regimes and for various spins $S\geq1/2$ have been discussed both numerically 
and analytically ~\cite{Kolezhuk2000,Lecheminant2001, Vekua2003, Hikihara2000, Hikihara2001, Hikihara2002,Kolezhuk2011} in 
the ferromagnetic~\cite{Hikihara2008} as well as antiferromagnetic regime~\cite{Hikihara2004, Furukawa2010,Azimi2014helical}.

Recent experiments on ultracold quantum gases in optical lattices~\cite{Aidelsburger2011, Struck2012, Miyake2013, Aidelsburger2013}, as well as irradiated graphene~\cite{oka2009,wang2013} or photonic lattices~\cite{Hafezi2011, Rechtsman2013, Mittal2016}, have paved the path towards the manipulation of 
lattice frustration to establish a situation to mimic condensed matter phenomena. The seminal experimental emulation of geometric frustration in a triangular optical lattice by Struck et al.~\cite{Struck2012} has attracted enormous interest to understand the physics of lattice frustration at ultra low temperature. 
In recent years various interesting predictions have been made in the context of geometric frustration in low dimensional lattices
such as zig-zag lattices which resembles the quantum $J_1$-$J_2$ model under proper conditions: 
Studies on systems of bosons in frustrated zig-zag lattices have predicted the presence of chiral phases~\cite{Greschner2013} 
which arise due the spontaneously breaking of the inversion symmetry of the system. On the other hand it has been shown 
that the supersolid phases can be stabilized in a system of hardcore bosons in a frustrated zig-zag lattice with dipole-dipole 
interactions~\cite{Mishra2014,Mishra2015molecule}. Recently, interesting extensions to an arbitrary rectified flux have been discussed~\cite{Anisimovas2016}.

A natural extension of the zig-zag ladder is to allow for a difference in the tunnelling amplitudes between upper and lower leg.
One of the interesting variant of the frustrated zig-zag lattice model is the sawtooth model
which exhibits non-trivial physics due to the existence of a flat band. It has been shown that a solid order emerges at quarter filling in a frustrated one dimensional sawtooth
model by Huber and Altman~\cite{Huber2010} by means of an effective model valid in the flat-band regime. 
Interestingly, a numerical analysis of this model has shown that also a supersolid phase can be stabilized in the absence of long-range interactions~\cite{Mishra2015sawtooth}. The existence of this supersolid phase can be attributed to the presence of alternating flux in the consecutive plaquettes of the lattice which occurs due the lattice geometry.

\begin{figure}[b]
\begin{center}
\includegraphics[width=\columnwidth]{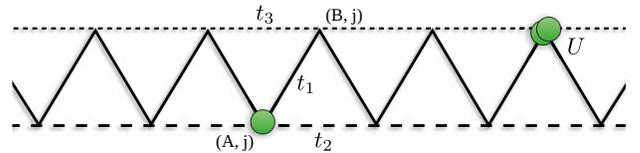}
 \end{center}
\caption{Railroad Trestle~(RRT) lattice which is a most general model for zig-zag ladder with tunneling amplitudes $t_1$, $t_2$ and $t_3$. }
\label{fig:rrtladder}
\end{figure}

In this paper we widen the scope of study to the general railroad-trestle~(RRT) model where one considers different 
hopping amplitudes in the legs of the ladder as shown in the Fig~\ref{fig:rrtladder}. The RRT model and its
variant the sawtooth model have been extensively analyzed in the context of 
fermions~\cite{Tonegawa1987,Sarkarrrt2002,Capriotti2003,Nakane2006,Sensawtooth1996}, but the bosonic or spin 
analog of this model is still a open problem. In this paper we present a detailed analysis of the ground-state 
properties of the bosonic RRT model in different limits 
to understand the effects of geometric frustration. We study three major variants of the RRT model using different analytical arguments in the 
limiting cases. The exact ground state properties are studied using the density matrix renormalization group~(DMRG) method~\cite{White1992, Schollwoeck2011}.

\section{Model}

The RRT model as sketched in Fig.~\ref{fig:rrtladder} is defined by the following Hamiltonian
\begin{align}
H(t_1,t_2,t_3) =& -t_1 \sum_{i} (a_{i}^{\dagger}b_{i}^{\phantom \dagger} + b_{i}^{\dagger}a_{i+1}^{\phantom \dagger} + \text{H.c.})\nonumber\\
&- t_2\sum_{i}(a_{i}^{\dagger}a_{i+1}^{\phantom \dagger}+\text{H.c.}) \nonumber\\
&- t_3\sum_{i}(b_{i}^{\dagger}b_{i+1}^{\phantom \dagger}+\text{H.c.})
\label{eq:rrtham}
\end{align}
Here, $a_i^{(\dagger)}$ and $b_i^{(\dagger)}$ are the bosonic annihilation(creation) operators for the upper (B) and lower (A) legs respectively (see Fig.~\ref{fig:rrtladder}). While $t_1$ is the hopping amplitude between the legs, $t_2$ and 
$t_3$ correspond to the hoppings along the leg-A and leg-B respectively. 
The local onsite interactions can be introduced in the model as 
\begin{align}
H_{int} = \frac{U}{2}\sum_{\nu\in\{A,B\},i}  n^\nu_i(n^\nu_i-1),
\label{eq:Hint}
\end{align}
where $U$ is the onsite repulsion and $n^\nu_i$ stands for the number operators at sites.  
In the following we assume the energy unit $t_1=1$ (unless stated otherwise) making all other physical quantities dimensionless. The primary focus of this work is to study the 
ground state properties of the Model~\eqref{eq:rrtham} 
in the limit of hardcore bosons ($U\to\infty$) for different values of $t_2$ and $t_3$ considering the 
frustrated regime i.e. $t_2<0$. It is now useful to introduce a dimensionless parameter 
\begin{align}
\delta=t_3/t_2 \,.
\end{align}

The remaining part of the paper is organized as follows. In the subsections of this section we analyze two limiting cases of the Model~\eqref{eq:rrtham}
such as the single particle spectrum and the limit of two decoupled chains i.e. when $|t_1| \ll |t2|,~|t3|$. In the following sections we discuss three different families of parameters: Section~(III) is devoted for the fully frustrated RRT(FF-RRT) 
model with $\pi$-$\pi$ flux arrangements, i.e. $t_3<0$. Sec.~(IV) constitutes the discussion on the $\pi$-$0$ flux case, with $t_3>0$. 
In Sec.~(V) we analyze the sawtooth ladder model i.e. $t_3=0$. In the end we conclude in Sec.~(IV).  

\begin{figure*}[t]
\begin{center}
\includegraphics[width=0.32\linewidth]{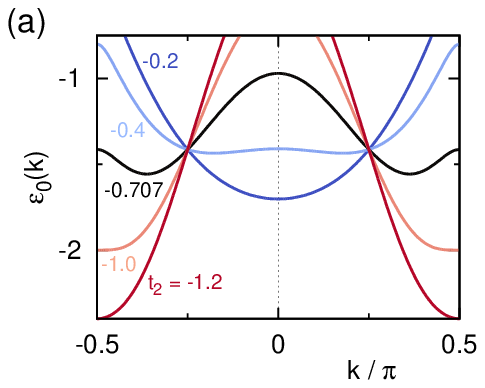}
\includegraphics[width=0.32\linewidth]{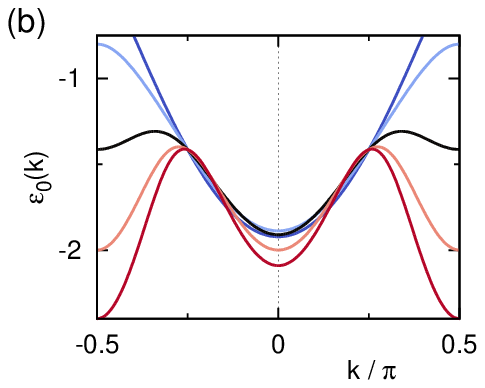}
\includegraphics[width=0.32\linewidth]{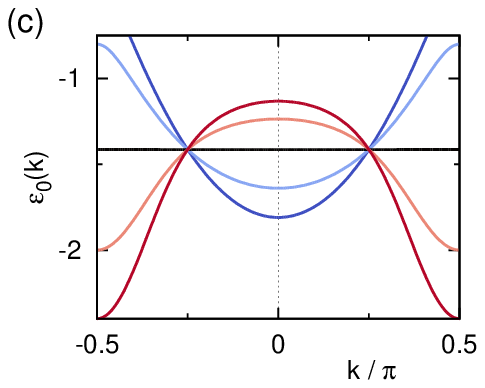}
\end{center}
\caption{Single particle energies of the RRT model with (a) $\pi$-$\pi$-case $\delta=t_2/t_3=1/2$, (b) $0-\pi$ case $\delta=-1/2$, and (c) sawtooth case $\delta=0$. 
We choose $t_2=-0.2$, $-0.4$, $-0.707$,$-1.0$, $-1.2$ (from top to bottom at $k\to\pi/2$).
\label{fig:disp}}
\end{figure*}

\subsection{Single particle spectrum}
It is instructive to start the discussion of the physics of Model~\eqref{eq:rrtham} from the single particle perspective. 
The kinetic part can be written in momentum space $k$ as
\begin{align}
H=-\sum_k \left(\!\!\begin{array}{c}
a_k\\
b_k\\
\end{array}\!\!\right)^\dagger
\!\left(
\begin{array}{cc}
 2 t_2 \cos (k)	 &	 t_1 \left(1+e^{i k}\right) \\
 t_1 \left(1+e^{-i k}\right)	 & 	2 t_3 \cos (k) \\
\end{array}
\right)
\left(\!\!\begin{array}{c}
a_k\\
b_k\\
\end{array}\!\!\right)
\end{align}
Diagonalizing the $2\times 2$ matrix one obtains the energy dispersion for generally two bands  as 
\begin{align}
\varepsilon_{0,1}(k) = \pm \sqrt{4 t_1^2 \cos^2\left(\frac{k}{2}\right) +  (t_2-t_3)^2 \cos^2(k)} \nonumber \\
-  (t_2+t_3) \cos(k)
\label{eq:singparten}
\end{align}
with the new creation and annihilation operators $\alpha_k=\cos(\theta_k) a_k +\sin(\theta_k) a_k $ and $\beta_k=\sin(\theta_k) a_k -\cos(\theta_k) a_k $, 
with the corresponding Bolgoliubov coefficients $\theta_k$. This expression for $\varepsilon_{0}(k)$ can give us insight into the physics of the system.

In general we are interested in three different cases, distinguished by the parameter $\delta=t_3/t_2$ (setting $t_2<0$). In Fig.~\ref{fig:disp} we 
show examples of the lowest band $\varepsilon_{0}(k)$ dispersion for
three different cases of $\delta$ and for each case we consider different values of $t_2$. For $\delta>0$, the flux through every unit-cell is 
equal to $\pi$ (Fig.~\ref{fig:disp}~(a)). Here one finds a parameter regime in which the dispersion exhibits a doubly degenerate minimum. 
For the case $\delta=1$ this model corresponds to the symmetric zig-zag ladder $H_S=H(t_1,t_2,t_2)$ resembling the $J_1-J_2$ model, 
which has been studied extensively in the literature as discussed in the introduction. 
In this case, the $\varepsilon_0(k)$ possesses single and double degenerate minima as a function of $t_2$ and 
becomes quartic ($\sim (k-Q)^4$) at the so called Lifshitz-transition point, $t_2=-1/4$. 

While for small values of $-t_2\ll 1$ the single minimum of the dispersion relation is at $k=0$, 
for large values of $-t_2\gg 1$ and $\delta\neq 1$ the dispersion relation will generally exhibit a 
minimum at $k=\pi/2$. We will later on associate two different single component Luttinger-liquid phases 
with these two dispersion minima, the superfluid at $k=0$ which we call the SF$_0$ phase, and the corresponding SF$_{\pi/2}$ phase at $k=\pi/2$. 
The situations in which the dispersion exhibits a degenerate minimum will give rise to further interesting quantum phases  discussed below in detail.

On the other hand, for $\delta<0$, only every second plaquette exhibits a $\pi$ flux while the others have zero flux. In this case, 
instead of a Lifshitz transition with a quartic dispersion relation,  the single-particle spectrum becomes degenerate only at a special point $\delta=\delta_c$ 
as shown in Fig.~\ref{fig:disp}~(b). 
This is, however, sufficient to induce a number of interesting effects in the ground state phase diagram which we will discuss later on.

Finally, for the special case of $\delta=0$ the system is called a sawtooth ladder. This exhibits a flat lowest band at $t_2=-1/\sqrt 2$ 
as shown in Fig.~\ref{fig:disp}~(c). 
Although apparently the sawtooth limit is the intermediate between the previous two cases i.e. $\delta<0$ and $\delta>0$, 
this situation resembles to some extent the $\pi$-$0$-flux systems 
as one bond is absent~\cite{Mishra2015sawtooth}. The many-body physics which translates from this kind of 
band picture will be systematically discussed in the following sections.

\subsection{Limit of decoupled chains $|t_1|\ll |t_2|,|t_3|$}

The phase diagram in the frustrated regime can be understood from the limit of two decoupled chains which is $|t_1| \ll |t_2|,|t_3|$ 
or in other words when $t_1\to 0$ the two chains are independent. 
For an asymmetric system, i.e. if $t_2\neq t_3$, both chains will in general be occupied by different particle densities. 
In the decoupling limit we expect only one chain with the larger tunneling amplitude $t_3>t_2$ to be occupied, 
if the density $n$ is small enough. This can be seen from a mapping to free fermions, 
which results in two bands $-2 t_2 \cos(k)$ and $-2 t_3 \cos(k)$. Only the lowest band is occupied for 
\begin{equation}
 n<\arccos\bigg(\frac{t_2}{t_3}\bigg)/2\pi .
 \label{eqn:density}
\end{equation}
For larger fillings the system enters a regime with two critical Luttinger liquids or two-superfluids (2SF), 
characterized by a central charge $c=2$~\cite{Calabrese2004}.

The effect of a perturbative coupling between the two chains i.e. by adding a small zig-zag hopping $H_{zz}=H(t_1,0,0)$  is best described by
a bosonization treatment of this case as presented in Ref.~\cite{Lecheminant2001} for the symmetric case $\delta=1$. For each sub-chain we introduce two pairs of 
bosonic fields ($\theta_{1},\phi_{1}$) and ($\theta_{2},\phi_{2}$). After forming symmetric and anti-symmetric combinations 
$\theta_\pm=(\theta_1\pm \theta_2)/\sqrt{2\pi}$, $\phi_\pm=\sqrt{\pi}(\phi_1\pm \phi_2)/\sqrt{2}$ the effective low-energy model~\cite{Greschner2013}  is given by 
\begin{align}
H = \sum_{\alpha=\pm}\frac{ v_{\alpha}}{2}\left[ \frac{(\partial_x \phi_{\alpha})^2}{ K_{\alpha}}+ K_{\alpha}(\partial_x \theta_{\alpha})^2  \right] \nonumber \\
+ \lambda\partial_x \theta_+\sin \sqrt{2\pi}\theta_-  + \cdots \,.
\end{align}
The last term is relevant and introduces a gap in the anti-symmetric sector $\theta_-$, resulting in a finite chirality $O_\chi \sim \langle \sin \sqrt{2\pi}\theta_-\rangle$~\cite{Nersesyan1998}. 
In the thermodynamic limit it exhibits a non-vanishing local boson current or chirality $\kappa_i=\frac{i}{2}(b_{i}^\dag b_{i+1}-{\rm H.c.})$ in the system which is a signature of the chiral superfluid(CSF) phase. 
In a finite system this locally defined chirality is always zero. However, the CSF phase is clearly characterized by the long-range 
ordered chirality-chirality correlations defined as
\begin{align}
O_\chi = \lim_{|i-j|\to\infty} \la{\kappa_i\kappa_j} \ra.
\end{align}
It is to be noted that the CSF phase possess a central charge $c=1$ and the 2SF phase does not exhibit a finite chirality.

Interestingly, for the anti-symmetric zig-zag model $H_A=H(t_1,t_2,-t_2)$ i.e. with $\delta=-1$, we do not expect this gapping mechanism to work. 
This can be understood by a simple gauge transformation 
$a_j^{(\dagger)} \to (-1)^{j} a_j^{(\dagger)}$, and $b_j^{(\dagger)} \to b_j^{(\dagger)}$. With this we can map $H_A\to H_S$, but the zig-zag hopping acquires an oscillating factor
\begin{align}
H_{zz} \to  \sum_{i} (-1)^{i} (a_{i}^{\dagger}b_{i}^{\phantom \dagger} - b_{i}^{\dagger}a_{i+1}^{\phantom \dagger} + \text{H.c.})
\end{align}
Due to this strong oscillatory term, the perturbation in general becomes irrelevant and the system should stay in the 2SF phase. Only for the case of a certain commensurability $n=1/4$, however, the oscillation may be compensated in a bosonization description, and we may expect the emergence of a gap in the symmetric sector. 

Note that the asymmetric case ($t_2\neq t_3$) may be understood as a combination of the symmetric and antisymmetric zig-zag model i.e. $H = \frac{t_2+t_3}{2} H_S + \frac{t_2-t_3}{2} H_A + H_{zz}$. 
Hence, we might naively expect the physics arising as a combination of both the effects. In the following we will examine these heuristic arguments  by means of more rigorous methods.

\section{The fully frustrated RRT~(FF-RRT) model ($\pi$-$\pi$-flux)}

In this section we begin the discussion with the case $\delta>0$ and then we compare our results with the already known case of the symmetric zig-zag chain. First we analyze the physics in the 
dilute limit and then we extend our calculation by increasing the density. 

\subsection{Dilute limit}

The interplay between local interactions and geometric frustration which gives rise to the various quantum phases can be best understood in the limit of 
low fillings $n\to 0$ or the dilute limit. In the presence of two non-equivalent minima at $k=\pm Q$ the ground state of a non-interacting boson system is
highly degenerate and the effect of interactions becomes crucial which selects a particular ground state. 
The particles at low energies mainly populate the two dispersion minima at $Q$ and $-Q$. We can interpret them as two different bosonic 
flavors and map to an effective two component model with intra-species coupling $g_{11}=g_{22}$ between bosons of the same species and 
inter-species coupling $g_{12}$ between different flavors. 
Typically two different types of SF ground states may be stabilized: Either the bosons equally occupy both minima, i.e. both 
flavors are present (the 2SF phase), or one of them is 
spontaneously selected and a one component SF phase with a spontaneously broken symmetry is realized.

If the intra-species coupling $g_{11}>g_{12}$, a two component Luttinger-liquid phase (2SF) may be realized. 
In this case both the dispersion minima are equally populated. 
On the other hand a dominant inter-species coupling $g_{11}<g_{12}$ results a spontaneously 
broken state with a dominant occupation of the dispersion minimum at $k=Q$ or $k=-Q$.

\begin{figure}[tb]
\begin{center}
\includegraphics[width=1\linewidth]{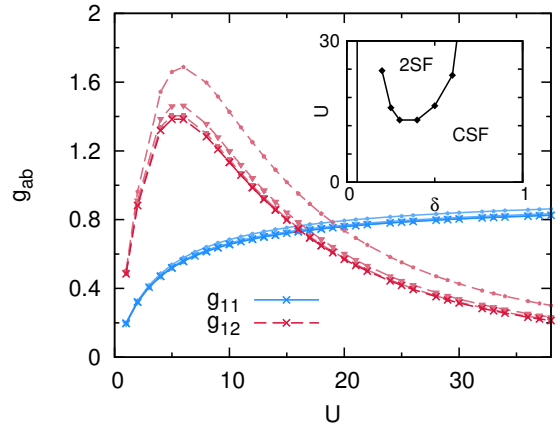}
\end{center}
\caption{Dilute limit intra-(inter-)particle coupling constants $g_{11}$($g_{12}$) are plotted w.r.t. $U$ for $t_2=-0.6$ and $\delta=0.5$. Different 
curves correspond to (top to bottom) different values of $E_* = 10^{-2}$, $10^{-3}$, and $10^{-4}$ and the cross-symbol denotes the extrapolation to $E_*\to0$. 
It can be seen that the value of $g_{11}$ dominates 
over $g_{12}$ after a critical value of $U$ indicating the 2SF phase. 
The inset shows the phase transition points between the dominant $g_{11}$ and $g_{12}$ corresponding to 
the 2SF and the CSF phases respectively as function of $U$ and $\delta$ for $t_2=-0.6$. The solid vertical line denotes the Lifsitz transition between SF$_0$ and frustrated phases. 
Close to this region our numerical scheme becomes unstable.
\label{fig:dilute}}
\end{figure}

While in general it is a useful observation~\cite{Kolezhuk2011} that both coupling coefficients, $g_{11}$ and $g_{12}$, may be extracted 
from the two particle scattering problem on the lattice, 
here we will follow a slightly different approach.
As shown in \cite{Kolezhuk2011} in the dilute limit it is possible to obtain the renormalized intra- and inter-component 
interactions analytically as an exact solution of the corresponding Bethe-Salpeter equation. A detailed analytical 
treatment can be found in Ref.~\cite{Kolezhuk2011}. 

For simplicity we will project the interaction to the lowest band. In momentum space the Hamiltonian becomes
\begin{equation}
H=\sum_k \epsilon(k) \beta^\alpha_k (\beta^\alpha_k)^\dagger + \frac{1}{2L} \sum_{k,k',q} V_q(k,k') \beta_{k+q} \beta_{k'-q} \beta_k \beta_{k'}
\end{equation}
where $V^{\alpha, \beta}_q(k,k')$ is the interaction between component $\alpha$ and $\beta$ in the momentum representation. 
For the BH model this is given by
\begin{align}
V_q(k_1,k_2)= \frac{U}{2} \left( \cos(\theta_{k_1})\cos(\theta_{k_2}) \cos(\theta_{k_2-q}) \cos(\theta_{k_1+q}) \right.\nonumber\\
\left. + \sin(\theta_{k_1}) \sin(\theta_{k_2}) \sin(\theta_{k_2-q}) \sin(\theta_{k_1+q}) \right)
\end{align}

We obtain the renormalized two-body interactions $\Gamma_{11}$ and $\Gamma_{12}$ 
in the dilute limit by the following form of the Bethe-Salpeter equations
\begin{align}
\Gamma_q^{11}(E) = V^{11}(Q,0) - \frac{1}{L} \sum_p \frac{V^{11}(q,p) \Gamma_p^{11}(E)}{\epsilon_{Q+p} + \epsilon_{Q-p} -E}
\label{eq:03_BS_gamma11}
\end{align}
and
\begin{align}
\Gamma_q^{12}(E) = 2 V^{12}(q,Q) - \frac{1}{L} \sum_p \frac{V^{12}(q,p) \Gamma_p^{12}(E)}{2\epsilon_{p}+E}
\label{eq:03_BS_gamma12}
\end{align}
where $E$ is the total energy of the incoming particles with momentum $k$ and $k'$. Here we have introduced the symmetriezed interactions
\begin{align}
V^{11}(q,p) &= \frac{1}{2} \left( V_{q-p}(Q+p, Q-p) + V_{q+p}(Q-p, Q+p) \right) \nonumber\\
V^{12}(q,p) &= \frac{1}{2} \left( V_{q-p}(-p, p) + V_{q+p}(-p, p) \right) \,.
\end{align}
$\Gamma_{11}$ and $\Gamma_{12}$ may be related to the bare coupling strengths $g_{11}$ and $g_{12}$ as 
\begin{align}
\frac{1}{\Gamma_{\alpha \beta}(-E_*)} = \left( \frac{m}{4 E_*} \right)^{1/2} + \frac{1}{g_{\alpha \beta}} + \mathcal{O}(E_*^{1/2}).
\label{eq:gamma_alpha_beta}
\end{align}
which corresponds to an off-shell regularization introducing a negative energy $E_*$. For $E_*\to 0$, corresponding 
to the dilute limit this procedure has been shown to be well defined. 
In the following we directly solve 
Eqs.~\eqref{eq:03_BS_gamma11} and ~\eqref{eq:03_BS_gamma12} numerically by introducing a 
Fourier representation of $\Gamma_q^{\alpha\beta}(-E_*)$ using a discretization of the equation and subsequent 
fast Fourier transform algorithm. The resulting linear set of equation 
can be solved using standard methods for finite values $E_*>0$ and subsequent extrapolation to $E_*\to 0$. 
This procedure becomes eventually unstable due to the presence of 
divergences in $\Gamma_q^{\alpha\beta}(E_*)$.

In Fig.~\ref{fig:dilute} we show the coupling constants as function of $U$ for the case $t_2/t_1=0.6$ and $\delta=0.5$. 
We extrapolate $g_{\alpha \beta}$ with a third order polynomial to the limit $E_*\to 0$. 
For weak interactions the inter-species couplings dominate. 
At a finite $U>U_c$ we observe a 
crossing between $g_{11}$ and $g_{12}$ curves and hence, a transition to the intra-species coupling dominated 2SF phase. 
In the inset of Fig.~\ref{fig:dilute} 
we show the extracted transition points $U=U_c$ as a function of $\delta$ for the case $t_2=-0.6$. It can be 
seen that as the value of $\delta$ increases the CSF phase becomes more robust and survives even in the large $U$ limit. 

\begin{figure}[t]
\begin{center}
\includegraphics[width=\columnwidth]{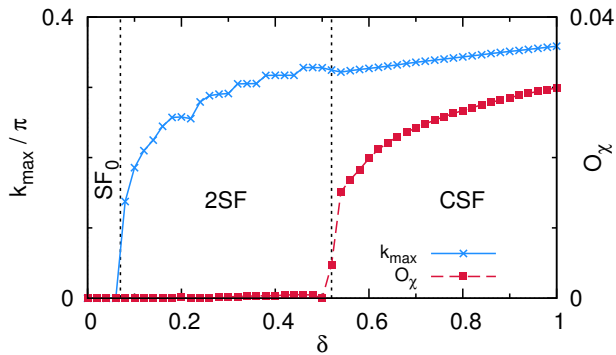}
\end{center}
\caption{Chirality $O_\chi$ and peak position $k_{max}$ of the momentum distribution $n(k)$ for the 
fully frustrated RRT model as function of $\delta>0$ for small fillings ($n=0.1$, $t_2=-0.6$, DMRG data, $L=80$). 
As $k_{max}\neq 0$ two equivalent maxima at $\pm k_{max}$ are found.
\label{fig:rrtzz_dilute_cc}}
\end{figure}

Now we perform numerical DMRG simulations to compare the results with the above findings for the example $t_2=-0.6$, also shown in Fig.~\ref{fig:dilute}. 
By considering a system of hardcore bosons with a finite but small filling $n=0.1$, 
we compute different order parameters such as the chirality order parameter $O_\chi$ and the momentum distribution function $n(k)$. 
The momentum distribution function is defined as 
\begin{align}
n(k)=\frac{1}{L^2}\sum_{i,j}{e^{ik(i-j)}} G_{ij}
\end{align}
with the single particle Greens functions $G_{ij}$ along the zig-zag direction of the chain.  
In Fig.~\ref{fig:rrtzz_dilute_cc} we plot both $O_\chi$ and the peak position of $n(k)$ as a function of $\delta$. 
One may clearly distinguish three regimes. For small values of $\delta$ there exists one peak in the momentum distribution function indicating an SF phase.
At some $\delta>\delta_{c1}$ the momentum distribution acquires a double peak structure with $k\neq 0$ which is a signature of the 2SF phase. 
For $\delta>\delta_{c2}$ the chirality becomes finite and the system enters into the CSF phase.

\begin{figure}[tb]
\begin{center}
\includegraphics[width=0.45\linewidth]{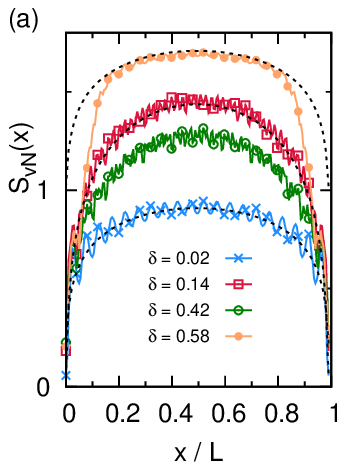}
\includegraphics[width=0.45\linewidth]{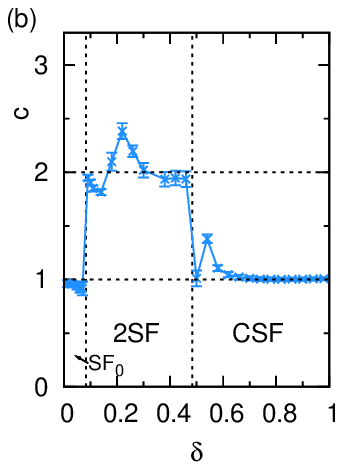}
\end{center}
\caption{Entanglement scaling for the RRT model using the same parameters as in Fig.~\ref{fig:rrtzz_dilute_cc} 
(DMRG data, $t_2=-0.6$, $L=201$ sites, filling $n=0.1$). (a) Entanglement entropy $S_{vN}(x)$ for different bi-partitions of 
the system $x$ for various values of $\delta$. The black dashed lines depict a fit to Eq.~\eqref{eq:ent}. 
(b) The extracted central charge $c$ from the fitting procedure.
\label{fig:ee_rrtzz}}
\end{figure}

Moreover, entanglement properties have been shown to provide useful general measure for the detection of 
quantum phase transitions~\cite{osterloh2002,Vidal2003}. In this regard, we calculate the von-Neumann entropy which is defined as 
\begin{equation}
S_{vN, L}(x) = -\mathrm{tr}\left( \rho_l \ln\rho_l \right) = \frac{c}{6} \ln\left[ \frac{L}{\pi} \sin\left(\frac{\pi}{L}l\right) \right] + g \,.
\label{eq:ent} 
\end{equation}
where, $\rho_l$ is the reduced density matrix for a subsystem of length $x$ which is plotted as function of $x/L$ in Fig.~\ref{fig:ee_rrtzz}(a). 
The right part of Eq.~\eqref{eq:ent} is valid for conformally invariant gapless states~\cite{Vidal2003, Calabrese2004}. 
We fit the expression in the r.h.s. of Eq.~\eqref{eq:ent} to the entanglement entropy curves obtained using the DMRG method as shown in Fig.~\ref{fig:ee_rrtzz}(a). From this we
extract the central charge $c$ of the underlying field-theory which is shown in Fig.~\ref{fig:ee_rrtzz}(b). 
Note, that for the RRT model we perform simulations of system sizes with odd number of sites in order 
to restore proper inversion symmetry at a central bond.
Consistent with our proceeding discussion in Fig.~\ref{fig:rrtzz_dilute_cc} we find that the 
intermediate non-chiral region exhibits an central charge $c=2$ and hence, can be called a 2SF phase. 

\begin{figure}[t]
\begin{center}
\includegraphics[width=1\linewidth]{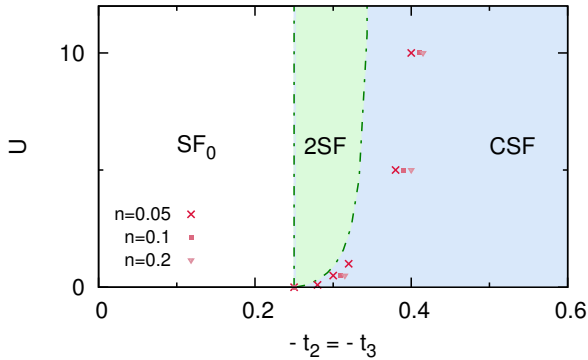}
\end{center}
\caption{Dilute limit phase diagram of interacting bosons in the symmetric zig-zag model ($\delta=1$). 
The data points show results for the 2SF to CSF transition from DMRG simulations at finite density $n>0$. 
\label{fig:dilute_zz}}
\end{figure}

For the special case of a symmetric zig-zag model $\delta=1$ we repeat this analysis in the dilute limit and using the DMRG method and obtain the 
phase diagram in the $U$-$t_2$-plane which is shown in Fig.~\ref{fig:dilute_zz}. 
Close to the Lifsitz transition the 2SF phase is realized. For large frustrations $|t_2|>1/\sqrt{8}$ no 
2SF phase is found and the system is in a CSF phase, which remains true for the hardcore bosons case. We compare our findings to 
DMRG results for various fillings and interaction strengths and, as shown in the figure, find a good qualitative 
agreement between the two results. The symbols in Fig.~\ref{fig:dilute_zz} shows the 2SF-CSF phase boundaries for different densities such as 
$n=0.05$~(cross), $n=0.1$~(squares) and $n=0.2$~(triangles). Note, that a direct comparison between the two methods may become difficult as 
for finite dilute systems the order parameter i.e. the chirality vanishes.

\subsection{Finite densities}

\begin{figure}[b]
\begin{center}
\includegraphics[width=1\linewidth]{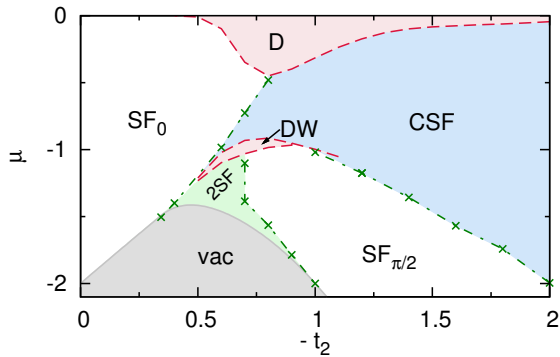}
\end{center}
\caption{Phase diagram of the FF RRT model for $\delta=1/2$ as function of $t_2 = 2 t_3$ and the chemical potential $\mu$.
\label{fig:pd_pipi}}
\end{figure}

In this subsection we will analyze the complete ground state phase diagram of the asymmetric FF-RRT model 
for a fixed $\delta=1/2$ as function of the chemical potential $\mu$ to understand the physics at finite densities. 
From the previous section we find that if $\delta=1/2$ 
for $\sqrt{\frac{3 \sqrt{33}}{2}-\frac{17}{2}} < - t_2 < 1$ the lowest band in Eq.~\eqref{eq:singparten} has a two fold degenerate 
minimum at $Q=\pm \frac{t_1 \left(3 \sqrt{2 t_1^2-t_2^2}-4 t_1\right)}{t_2^2}$. We explore the physics of this system for different values of $t_2$ by varying the chemical 
potential $\mu$. 
\begin{figure}[t]
\begin{center}
\includegraphics[width=1\linewidth]{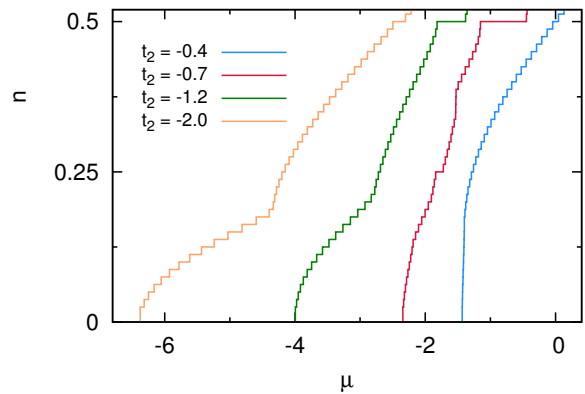}
\end{center}
\caption{$\mu$-$n$-curve for cuts through phase diagram Fig.~\ref{fig:pd_pi0} for $\delta=1/2$ and (from left to right) 
$t_2=-2.0$, $-1.2$, $-1.0$ and $-0.7$.
\label{fig:mag_pipi}}
\end{figure}
In Fig.~\ref{fig:pd_pipi} we show the phase diagram in the $\mu$-$t_2$-plane. Consistent with the proceeding section we do not find 
the emergence of a CSF phase at small values of $\delta$ in the dilute limit. However, at larger fillings the system 
enters an extensive CSF region. Apart from this, other interesting features appear in the phase diagram which we discuss below. 

The phase transition points can be best read from the $\mu$-$n$-diagrams of finite systems which is shown in Fig.~\ref{fig:mag_pipi} for different values of $t_2$. 
At the transition points between the single component superfluid phases such as the SF and the SF$_{\pi/2}$ phases 
and the CSF or 2SF phases the $\mu$-$n$-curve exhibits a sharp kink.
In order to distinguish the 2SF and CSF phases we use the the chirality order parameter and the central charge as discussed before. 
We observe the SF$_{\pi/2}$-CSF transition for a critical density $n_c \approx 0.18$ (for $t_2=-2$) which is consistent with $n_c=1/6$ that is already 
obtained in the decoupled chain limit using Eq.~\eqref{eqn:density}. 

\begin{figure}[tb]
\begin{center}
\includegraphics[height=5.9cm]{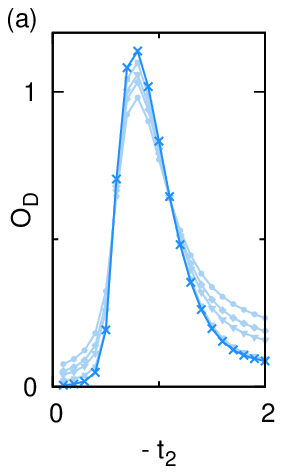}
\includegraphics[height=5.9cm]{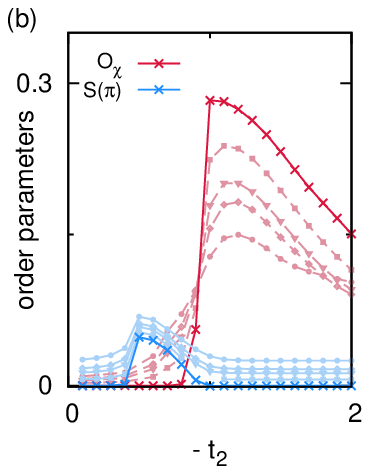}
\end{center}
\caption{Order parameters for different cuts through the phase diagram for (a) $n=1/2$ and (b) $n=1/4$. 
The curves in lighter shadings show the finite-size results for $L=20$ (circle), $40$  (diamond), $80$  (triangle) and $160$ (box) sites - cross-symbols
depict the extrapolation to the thermodynamic limit using a higher order polynomial.
\label{fig:cut_pipi}}
\end{figure}

The $\mu$-$n$-curves of Fig.~\ref{fig:mag_pipi} show a series of plateaus at certain commensurate fillings, $n=1/4$ and $n=1/2$. 
These correspond to the gaped insulating phases, a density wave(DW) phase(at $n=1/4$) and a dimerized(D) phase ($n=1/2$), which are stabilized due to 
frustration and asymmetry of the model. As discussed in Ref.~\cite{Greschner2013} at the Lifshitz transition, the band curvature 
vanishes locally as the minimum becomes quartic. Hence, as the effective mass diverges we may expect the pinning of particles at 
weak interaction strengths resulting into the emergence of gaped phases. In Fig.~\ref{fig:pd_pipi} we show the approximate extent of 
the plateau regions bounded by the dashed curves which are calculated for several finite system sizes 
and then extrapolated to the thermodynamic limit by means of a higher order polynomial. For the case of hardcore bosons, the presence of a D 
phase at half filling $n=1/2$ (for zero magnetic field in the case of the corresponding spin-1/2 model) has been discussed 
extensively~\cite{Okamoto1992, Hikihara2001}.
Following Okamoto and Nomura~\cite{Okamoto1992} we may extract the phase transition points between the SF$_0$ and the D phase by means of 
a level crossing analysis. To further characterize the 
D phase we compute the dimer-dimer order parameter as 
\begin{equation}
O_D =\frac{1}{L}\sum_{i}(-1)^i B_i,
\end{equation}
where $B_i=\langle b_i (b_{i+1}^\dagger+b_{i-1}^\dagger)\rangle$ is the bond energy. 
In Fig.~\ref{fig:cut_pipi}~(a) we show the behaviour of $O_D$ at half filling as a function of $t_2$ for 
different system sizes $L=20,~40,~80,~160$ along with the extrapolated curve in the 
thermodynamic limit. 

Interestingly, for the RRT model we also find an emerging density wave (DW) phase at quarter filling $n=1/4$ close 
to the Lifshitz line. The emerging DW order can be seen as a peak in the density structure factor 
\begin{equation}
 S(k)=\frac{1}{L^2}\sum_{i,j}{e^{ik(i-j)}}\langle{n_{i}n_{j}}\rangle ,
\end{equation}
where $\langle{n_{i}n_{j}}\rangle$ is the density-density correlation between sites $i$ and $j$. In Fig.~\ref{fig:cut_pipi}~(b) we 
plot the values of $S(k=\pi)$(blue symbols) 
and the chirality $O_\chi$(red symbols) as a function of 
$t_2/t_1$ for different lengths and also in the thermodynamic limit at $n=1/4$. This clearly shows the presence of the DW phase for some intermediate range of $t_2$ and the system possesses finite 
chirality for larger values of $t_2$ where a CSF phase is found. Note that the chirality becomes finite abruptly with the vanishing of the DW-order parameter as we enter the CSF phase.

\subsection{Symmetric zig-zag model}

\begin{figure}[t]
\begin{center}
\includegraphics[width=1\linewidth]{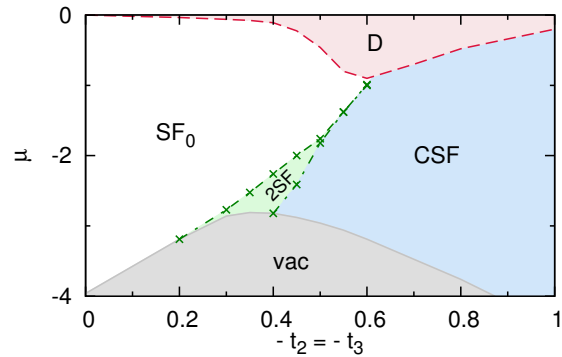}
\end{center}
\caption{Phase diagram for hardcore bosons in a symmetric frustrated zig-zag ladder~($\delta=1$) in the $t_2=t_3$ and $\mu$ plane.
\label{fig:pd_zz}}
\end{figure}

Contrary to the previously discussed case, for the symmetric zig-zag model~($\delta=1$), the dispersion relation is doubly 
degenerate for every $-t_2>1/4$. For completeness we depict the corresponding phase diagram in  Fig.~\ref{fig:pd_zz}.
Here, we find an extended CSF phase for any filling as $-t_2$ is large enough. For small densities, close to the 
Lifsitz transition the interesting interplay between the 2SF and CSF phases is observed. The transition point from the 
low density description is consistent with the numerical simulations.
Due to the symmetry of the model the DW phase at quarter filling is absent. However, there exists a D phase at $n=1/2$ as a result of frustration.

\section{The $\pi$-$0$ case}
\begin{figure}[t]
\begin{center}
\includegraphics[width=1\linewidth]{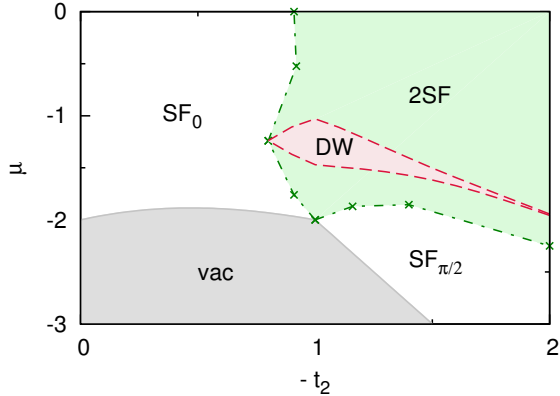}
\end{center}
\caption{Phase diagram of the $\pi$-$0$ RRT model with $\delta=-1/2$.
\label{fig:pd_pi0}}
\end{figure}

\begin{figure}[b]
\begin{center}
\includegraphics[width=1\linewidth]{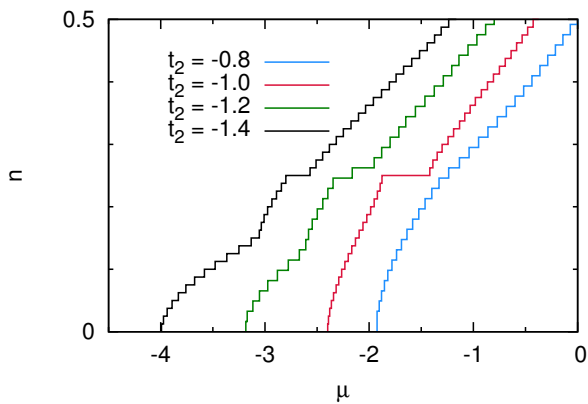}
\end{center}
\caption{$n$-$\mu$-curve for cuts through phase diagram of Fig.~\ref{fig:pd_pi0} for $\delta=-1/2$ 
and  $t_2 =-1.4$, $-1.2$, $-1.0$ and $-0.8$(from left to right).
\label{fig:mag_pi0}}
\end{figure}

Let us now turn to the anti-symmetric case when $\delta<0$, i.e. a model with a $\pi$ flux through every second plaquette. 
Here we analyze this model along the line discussed above and obtain the complete phase diagram as shown in Fig.~\ref{fig:pd_pi0} 
for $\delta=-0.5$. The phase diagram is 
obtained by analyzing the plateaus in the $\mu-n$ plot (Fig.~\ref{fig:mag_pi0}) 
and the order parameters as done in the previous case. Fig.~\ref{fig:mag_pi0} shows the 
emergence of plateaus only at $n=1/4$ which corresponds the the DW phase. This DW phase is denoted by the region bounded by the 
dashed curve in Fig.~\ref{fig:pd_pi0}. Interestingly a gapped phase at half filling is absent in this case. 
The extent of the DW phase is drastically enhanced compared to the case of a $\pi$-$\pi$-flux. 
In particular, for large values of $-t_2$ we still observe a finite gap after extrapolation of our numerical data to the thermodynamic limit. 
The grey region bounded by the continuous line is the empty state. 

\begin{figure}[t]
\begin{center}
\includegraphics[width=0.45\linewidth]{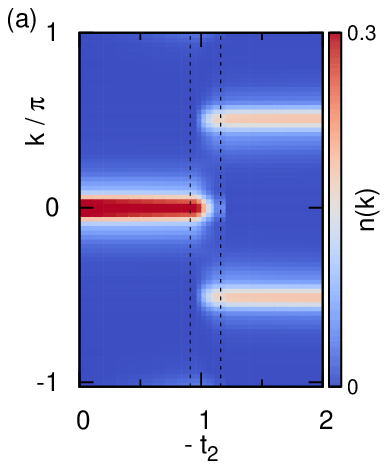}
\includegraphics[width=0.45\linewidth]{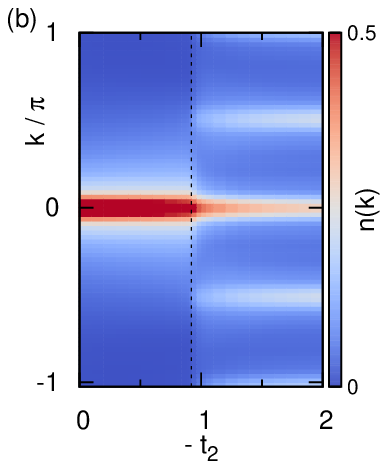}
\end{center}
\caption{Momentum distributions for filling $n=1/8$ and $3/8$ which corresponds to two cuts in the phase diagram of Fig.~\ref{fig:pd_pi0}.
\label{fig:mom_pi0}}
\end{figure}

As discussed in Sec.~(II), there should not exist a CSF phase in this scenario for weakly coupled chains, 
which we find to remain valid also for a  finite inter-leg hopping. We confirm this using our DMRG calculation and indeed, we see a broad 
region of the 2SF phase around the gapped DW phase marked by the dashed-cross boundary. The transition to the 2SF phase is characterized 
by a series of kinks in the $\mu$-$n$-curve~(see Fig.~\ref{fig:pd_pi0}). 

\begin{figure}[b]
\begin{center}
\includegraphics[width=0.45\linewidth]{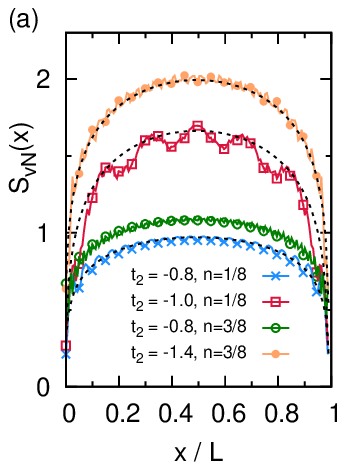}
\includegraphics[width=0.45\linewidth]{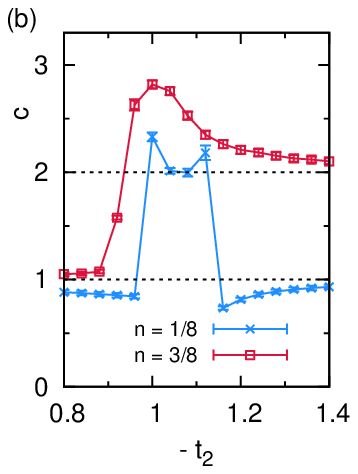}
\end{center}
\caption{Entanglement scaling for the $\delta<0$ RRT model using the same parameters as in 
Fig.~\ref{fig:rrtzz_dilute_cc} (DMRG data, $L=201$ sites, filling $n=0.1$). (a) Entanglement entropy $S_{vN}(x)$ for 
different bi-partitions of the system $x$ for various values of $\delta$. The black dashed lines depict a fit 
to Eq.~\eqref{eq:ent}. (b) The extracted central charge $c$ from the fitting procedure.
\label{fig:ee_pi0}}
\end{figure}

The SF$_0$ and SF$_{\pi/2}$ phases are best understood by looking at the momentum distribution function $n(k)$ as plotted in Fig.\ref{fig:mom_pi0}. 
We plot $n(k)$ for two cuts through the 
phase diagram of Fig.\ref{fig:pd_pi0} along the X-axis which correspond to two different fillings $n=1/8$ and $n=3/8$ in Fig.\ref{fig:mom_pi0}(a) and Fig.\ref{fig:mom_pi0}(b) respectively. 
For the cut along $n=1/8$, the momentum distribution exhibits one peak at $k_0$, 
then three peaks and in the end two peaks at $k=\pm\pi/2$ as a function of $t_2$. 
While the SF$_{\pi/2}$ phase is characterized by peaks at $k=\pm \pi/2$, which are equivalent, 
in the 2SF phase region we find multi-peak structure with peaks at $k=0$ and $\pm \pi/2$. This means the system goes from 
the SF to the SF$_{\pi/2}$ phase and then to the 2SF phase. 
In the case of $n=3/8$, there is a single transition from the SF to the 2SF phase as can be seen from Fig.\ref{fig:mom_pi0}(b). The phase transitions between 
these superfluid phases are marked by the vertical dashed lines in Fig.~\ref{fig:mom_pi0}. 
We also compute the central charge $c$ following the analysis done in the 
previous section and show that the numerical estimation of the central charge is consistent with $c=1$ in the SF$_0$ and SF$_{\pi/2}$ phases where as 
$c=2$ in the 2SF regions(see Fig.~\ref{fig:ee_pi0}).

\section{The sawtooth chain}
In the end we analyze the very special case of the RRT model which is known as the sawtooth chain. 
As stated in the introduction, for the sawtooth case ($\delta=0$) the lowest band becomes exactly 
flat at the special value of $t_2=-1/\sqrt{2}$ (see Fig.~\ref{fig:disp}(c)).
\begin{figure}[t]
\begin{center}
\includegraphics[width=1\linewidth]{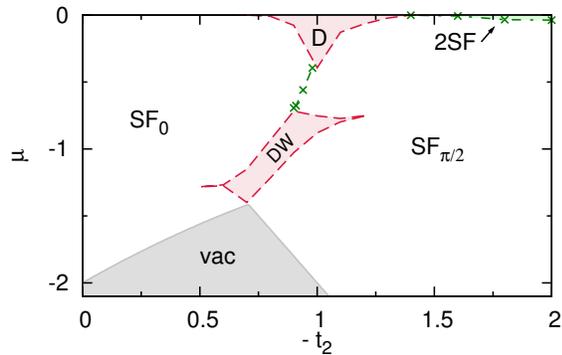}
\end{center}
\caption{The phase diagram for the sawtooth ladder model for hardcore bosons.
\label{fig:pd_saw}}
\end{figure}
\begin{figure}[b]
\begin{center}
\includegraphics[width=1\linewidth]{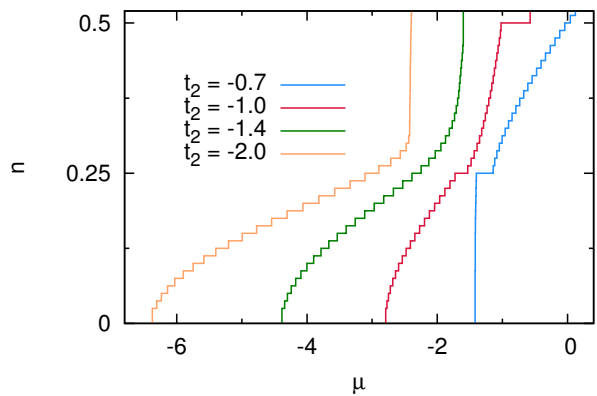}
\end{center}
\caption{$\mu$-$n$-curve for  $t_2 =-0.7$, $-1.0$, $-1.4$ and $-2.0$ (right to left) for the sawtooth model.
\label{fig:mag_saw}}
\end{figure}
Here we analyze the sawtooth model for the hardcore bosons case and obtain the interesting ground state phase 
diagram which is shown in Fig.~\ref{fig:pd_saw} . 
Examples of the equation of state from which the main results can be deduced are shown in Fig.~\ref{fig:mag_saw}.

The presence of the flat band leads, as for the Lifshitz transitions, to an enhancement of correlations. As a result we find an 
extensive D and DW phase around $t_2=-1/\sqrt{2}$ which are bounded by the dashed curves  in Fig.~\ref{fig:pd_saw}  at $n=1/2$ and $n=1/4$ respectively. 
The presence of the flat-band also leads to
macroscopically large jumps in density in the $\mu-n$ curve for fillings below $n=1/4$. The transition between the SF 
and SF$_{\pi/2}$ phase is apparently direct, possibly of first order. 
For the hardcore case we do not observe an emerging supersolid phase like the softcore case discussed in Ref.~\cite{Mishra2015sawtooth}, 
however, we find a 2SF phase for large 
fillings and $-t_2\gtrsim 1.4$. As seen in Fig.~\ref{fig:mag_saw} it is characterized by a sharp 
increase in the density which indicates a very large but finite compressibility.


\section{Summary}

In summary in this paper we have studied the ground-state physics of a very generic zig-zag ladder model, with asymmetric 
hopping strengths on the two legs. The interplay between this asymmetry and the interactions of the bosonic particles 
gives rise to various phenomena and quantum phases including the 2SF and the CSF phases and different single component SF phases. 
At certain commensurate fillings density wave and dimerized phases can be observed. While for the symmetric case chiral phases dominate 
the grand canonical phase diagram, the asymmetry tends to stabilize the 2SF phases.

In state of the art ultra-cold atom experiments the RRT models should in a natural way emerge from the attempts to study the symmetric 
zig-zag ladder models. For example one may realize a zig-zag model by means of superlattice techniques on triangular lattices in 
combination with lattice shaking~\cite{Struck2012,Greschner2013}. A slight misalignment of superlattice and the triangular lattice 
might typically lead to the tunneling asymmetry described here. Also one might adapt synthetic dimension approaches as recently 
proposed in Ref.\cite{Anisimovas2016}, where the requirement of a state-dependent lattice also may be naturally exploited to generalize to RRT-type models.

\begin{acknowledgments}
We would like to thank Luis Santos and Temo Vekua for important discussions.
S.G. acknowledges support of the German Research Foundation DFG (project no. SA 1031/10-1) and of the Swiss National Science 
Foundation under Division II. T.M. acknowledges hospitality of the Institute for Theoretical Physics Hannover, 
where part of this work has been carried out and also DST-SERB for the early career grant
through Project No. ECR/2017/001069. 
Simulations were carried out on the cluster system at the Leibniz University of Hannover, Germany.
\end{acknowledgments}

\bibliography{references}

\begin{thebibliography}{44}
\expandafter\ifx\csname natexlab\endcsname\relax\def\natexlab#1{#1}\fi
\expandafter\ifx\csname bibnamefont\endcsname\relax
  \def\bibnamefont#1{#1}\fi
\expandafter\ifx\csname bibfnamefont\endcsname\relax
  \def\bibfnamefont#1{#1}\fi
\expandafter\ifx\csname citenamefont\endcsname\relax
  \def\citenamefont#1{#1}\fi
\expandafter\ifx\csname url\endcsname\relax
  \def\url#1{\texttt{#1}}\fi
\expandafter\ifx\csname urlprefix\endcsname\relax\def\urlprefix{URL }\fi
\providecommand{\bibinfo}[2]{#2}
\providecommand{\eprint}[2][]{\url{#2}}

\bibitem[{\citenamefont{Hase et~al.}(2004)\citenamefont{Hase, Kuroe, Ozawa,
  Suzuki, Kitazawa, Kido, and Sekine}}]{Hase2004magnetic}
\bibinfo{author}{\bibfnamefont{M.}~\bibnamefont{Hase}},
  \bibinfo{author}{\bibfnamefont{H.}~\bibnamefont{Kuroe}},
  \bibinfo{author}{\bibfnamefont{K.}~\bibnamefont{Ozawa}},
  \bibinfo{author}{\bibfnamefont{O.}~\bibnamefont{Suzuki}},
  \bibinfo{author}{\bibfnamefont{H.}~\bibnamefont{Kitazawa}},
  \bibinfo{author}{\bibfnamefont{G.}~\bibnamefont{Kido}}, \bibnamefont{and}
  \bibinfo{author}{\bibfnamefont{T.}~\bibnamefont{Sekine}},
  \bibinfo{journal}{Phys. Rev. B} \textbf{\bibinfo{volume}{70}},
  \bibinfo{pages}{104426} (\bibinfo{year}{2004}).

\bibitem[{\citenamefont{Masuda et~al.}(2005)\citenamefont{Masuda, Zheludev,
  Roessli, Bush, Markina, and Vasiliev}}]{Masuda2005spin}
\bibinfo{author}{\bibfnamefont{T.}~\bibnamefont{Masuda}},
  \bibinfo{author}{\bibfnamefont{A.}~\bibnamefont{Zheludev}},
  \bibinfo{author}{\bibfnamefont{B.}~\bibnamefont{Roessli}},
  \bibinfo{author}{\bibfnamefont{A.}~\bibnamefont{Bush}},
  \bibinfo{author}{\bibfnamefont{M.}~\bibnamefont{Markina}}, \bibnamefont{and}
  \bibinfo{author}{\bibfnamefont{A.}~\bibnamefont{Vasiliev}},
  \bibinfo{journal}{Phys. Rev. B} \textbf{\bibinfo{volume}{72}},
  \bibinfo{pages}{014405} (\bibinfo{year}{2005}).

\bibitem[{\citenamefont{Drechsler et~al.}(2007)\citenamefont{Drechsler,
  Volkova, Vasiliev, Tristan, Richter, Schmitt, Rosner, M{\'a}lek, Klingeler,
  Zvyagin et~al.}}]{Drechsler2007frustrated}
\bibinfo{author}{\bibfnamefont{S.-L.} \bibnamefont{Drechsler}},
  \bibinfo{author}{\bibfnamefont{O.}~\bibnamefont{Volkova}},
  \bibinfo{author}{\bibfnamefont{A.}~\bibnamefont{Vasiliev}},
  \bibinfo{author}{\bibfnamefont{N.}~\bibnamefont{Tristan}},
  \bibinfo{author}{\bibfnamefont{J.}~\bibnamefont{Richter}},
  \bibinfo{author}{\bibfnamefont{M.}~\bibnamefont{Schmitt}},
  \bibinfo{author}{\bibfnamefont{H.}~\bibnamefont{Rosner}},
  \bibinfo{author}{\bibfnamefont{J.}~\bibnamefont{M{\'a}lek}},
  \bibinfo{author}{\bibfnamefont{R.}~\bibnamefont{Klingeler}},
  \bibinfo{author}{\bibfnamefont{A.}~\bibnamefont{Zvyagin}},
  \bibnamefont{et~al.}, \bibinfo{journal}{Phys. Rev. Lett.}
  \textbf{\bibinfo{volume}{98}}, \bibinfo{pages}{077202}
  (\bibinfo{year}{2007}).

\bibitem[{\citenamefont{Vasiliev et~al.}(2018)\citenamefont{Vasiliev, Volkova,
  Zvereva, and Markina}}]{vasiliev2018milestones}
\bibinfo{author}{\bibfnamefont{A.}~\bibnamefont{Vasiliev}},
  \bibinfo{author}{\bibfnamefont{O.}~\bibnamefont{Volkova}},
  \bibinfo{author}{\bibfnamefont{E.}~\bibnamefont{Zvereva}}, \bibnamefont{and}
  \bibinfo{author}{\bibfnamefont{M.}~\bibnamefont{Markina}},
  \bibinfo{journal}{npj Quantum Materials} \textbf{\bibinfo{volume}{3}},
  \bibinfo{pages}{18} (\bibinfo{year}{2018}).

\bibitem[{\citenamefont{Majumdar and Ghosh}(1969)}]{Majumdar1969}
\bibinfo{author}{\bibfnamefont{C.~K.} \bibnamefont{Majumdar}} \bibnamefont{and}
  \bibinfo{author}{\bibfnamefont{D.~K.} \bibnamefont{Ghosh}},
  \bibinfo{journal}{Journal of Mathematical Physics}
  \textbf{\bibinfo{volume}{10}}, \bibinfo{pages}{1388} (\bibinfo{year}{1969}).

\bibitem[{\citenamefont{Haldane}(1982)}]{Haldane1982}
\bibinfo{author}{\bibfnamefont{F.}~\bibnamefont{Haldane}},
  \bibinfo{journal}{Phys. Rev. B} \textbf{\bibinfo{volume}{25}},
  \bibinfo{pages}{4925} (\bibinfo{year}{1982}).

\bibitem[{\citenamefont{Okamoto and Nomura}(1992)}]{Okamoto1992}
\bibinfo{author}{\bibfnamefont{K.}~\bibnamefont{Okamoto}} \bibnamefont{and}
  \bibinfo{author}{\bibfnamefont{K.}~\bibnamefont{Nomura}},
  \bibinfo{journal}{Phys. Lett. A} \textbf{\bibinfo{volume}{169}},
  \bibinfo{pages}{433} (\bibinfo{year}{1992}).

\bibitem[{\citenamefont{Kolezhuk}(2000)}]{Kolezhuk2000}
\bibinfo{author}{\bibfnamefont{A.~K.} \bibnamefont{Kolezhuk}},
  \bibinfo{journal}{Phys. Rev. B} \textbf{\bibinfo{volume}{62}},
  \bibinfo{pages}{R6057} (\bibinfo{year}{2000}).

\bibitem[{\citenamefont{Lecheminant et~al.}(2001)\citenamefont{Lecheminant,
  Jolicoeur, and Azaria}}]{Lecheminant2001}
\bibinfo{author}{\bibfnamefont{P.}~\bibnamefont{Lecheminant}},
  \bibinfo{author}{\bibfnamefont{T.}~\bibnamefont{Jolicoeur}},
  \bibnamefont{and} \bibinfo{author}{\bibfnamefont{P.}~\bibnamefont{Azaria}},
  \bibinfo{journal}{Phys. Rev. B} \textbf{\bibinfo{volume}{63}},
  \bibinfo{pages}{174426} (\bibinfo{year}{2001}).

\bibitem[{\citenamefont{Vekua et~al.}(2003)\citenamefont{Vekua, Japaridze, and
  Mikeska}}]{Vekua2003}
\bibinfo{author}{\bibfnamefont{T.}~\bibnamefont{Vekua}},
  \bibinfo{author}{\bibfnamefont{G.}~\bibnamefont{Japaridze}},
  \bibnamefont{and} \bibinfo{author}{\bibfnamefont{H.-J.}
  \bibnamefont{Mikeska}}, \bibinfo{journal}{Phys. Rev. B}
  \textbf{\bibinfo{volume}{67}}, \bibinfo{pages}{064419}
  (\bibinfo{year}{2003}).

\bibitem[{\citenamefont{Hikihara et~al.}(2000)\citenamefont{Hikihara, Kaburagi,
  Kawamura, and Tonegawa}}]{Hikihara2000}
\bibinfo{author}{\bibfnamefont{T.}~\bibnamefont{Hikihara}},
  \bibinfo{author}{\bibfnamefont{M.}~\bibnamefont{Kaburagi}},
  \bibinfo{author}{\bibfnamefont{H.}~\bibnamefont{Kawamura}}, \bibnamefont{and}
  \bibinfo{author}{\bibfnamefont{T.}~\bibnamefont{Tonegawa}},
  \bibinfo{journal}{J. Phys. Soc. Jpn.} \textbf{\bibinfo{volume}{69}}
  (\bibinfo{year}{2000}).

\bibitem[{\citenamefont{Hikihara et~al.}(2001)\citenamefont{Hikihara, Kaburagi,
  and Kawamura}}]{Hikihara2001}
\bibinfo{author}{\bibfnamefont{T.}~\bibnamefont{Hikihara}},
  \bibinfo{author}{\bibfnamefont{M.}~\bibnamefont{Kaburagi}}, \bibnamefont{and}
  \bibinfo{author}{\bibfnamefont{H.}~\bibnamefont{Kawamura}},
  \bibinfo{journal}{Phys. Rev. B} \textbf{\bibinfo{volume}{63}},
  \bibinfo{pages}{174430} (\bibinfo{year}{2001}).

\bibitem[{\citenamefont{Hikihara}(2002)}]{Hikihara2002}
\bibinfo{author}{\bibfnamefont{T.}~\bibnamefont{Hikihara}},
  \bibinfo{journal}{J. Phys. Soc. Jpn.} \textbf{\bibinfo{volume}{71}},
  \bibinfo{pages}{319} (\bibinfo{year}{2002}).

\bibitem[{\citenamefont{Kolezhuk et~al.}(2012)\citenamefont{Kolezhuk,
  Heidrich-Meisner, Greschner, and Vekua}}]{Kolezhuk2011}
\bibinfo{author}{\bibfnamefont{A.}~\bibnamefont{Kolezhuk}},
  \bibinfo{author}{\bibfnamefont{F.}~\bibnamefont{Heidrich-Meisner}},
  \bibinfo{author}{\bibfnamefont{S.}~\bibnamefont{Greschner}},
  \bibnamefont{and} \bibinfo{author}{\bibfnamefont{T.}~\bibnamefont{Vekua}},
  \bibinfo{journal}{Phys. Rev. B} \textbf{\bibinfo{volume}{85}},
  \bibinfo{pages}{064420} (\bibinfo{year}{2012}).

\bibitem[{\citenamefont{Hikihara et~al.}(2008)\citenamefont{Hikihara, Kecke,
  Momoi, and Furusaki}}]{Hikihara2008}
\bibinfo{author}{\bibfnamefont{T.}~\bibnamefont{Hikihara}},
  \bibinfo{author}{\bibfnamefont{L.}~\bibnamefont{Kecke}},
  \bibinfo{author}{\bibfnamefont{T.}~\bibnamefont{Momoi}}, \bibnamefont{and}
  \bibinfo{author}{\bibfnamefont{A.}~\bibnamefont{Furusaki}},
  \bibinfo{journal}{Phys. Rev. B} \textbf{\bibinfo{volume}{78}},
  \bibinfo{pages}{144404} (\bibinfo{year}{2008}).

\bibitem[{\citenamefont{Hikihara and Furusaki}(2004)}]{Hikihara2004}
\bibinfo{author}{\bibfnamefont{T.}~\bibnamefont{Hikihara}} \bibnamefont{and}
  \bibinfo{author}{\bibfnamefont{A.}~\bibnamefont{Furusaki}},
  \bibinfo{journal}{Phys. Rev. B} \textbf{\bibinfo{volume}{69}},
  \bibinfo{pages}{064427} (\bibinfo{year}{2004}).

\bibitem[{\citenamefont{Furukawa et~al.}(2010)\citenamefont{Furukawa, Sato, and
  Onoda}}]{Furukawa2010}
\bibinfo{author}{\bibfnamefont{S.}~\bibnamefont{Furukawa}},
  \bibinfo{author}{\bibfnamefont{M.}~\bibnamefont{Sato}}, \bibnamefont{and}
  \bibinfo{author}{\bibfnamefont{S.}~\bibnamefont{Onoda}},
  \bibinfo{journal}{Phys. Rev. Lett.} \textbf{\bibinfo{volume}{105}},
  \bibinfo{pages}{257205} (\bibinfo{year}{2010}).

\bibitem[{\citenamefont{Azimi et~al.}(2014)\citenamefont{Azimi, Chotorlishvili,
  Mishra, Greschner, Vekua, and Berakdar}}]{Azimi2014helical}
\bibinfo{author}{\bibfnamefont{M.}~\bibnamefont{Azimi}},
  \bibinfo{author}{\bibfnamefont{L.}~\bibnamefont{Chotorlishvili}},
  \bibinfo{author}{\bibfnamefont{S.}~\bibnamefont{Mishra}},
  \bibinfo{author}{\bibfnamefont{S.}~\bibnamefont{Greschner}},
  \bibinfo{author}{\bibfnamefont{T.}~\bibnamefont{Vekua}}, \bibnamefont{and}
  \bibinfo{author}{\bibfnamefont{J.}~\bibnamefont{Berakdar}},
  \bibinfo{journal}{Phys. Rev. B} \textbf{\bibinfo{volume}{89}},
  \bibinfo{pages}{024424} (\bibinfo{year}{2014}).

\bibitem[{\citenamefont{Aidelsburger et~al.}(2011)\citenamefont{Aidelsburger,
  Atala, Nascimb\`ene, Trotzky, Chen, and Bloch}}]{Aidelsburger2011}
\bibinfo{author}{\bibfnamefont{M.}~\bibnamefont{Aidelsburger}},
  \bibinfo{author}{\bibfnamefont{M.}~\bibnamefont{Atala}},
  \bibinfo{author}{\bibfnamefont{S.}~\bibnamefont{Nascimb\`ene}},
  \bibinfo{author}{\bibfnamefont{S.}~\bibnamefont{Trotzky}},
  \bibinfo{author}{\bibfnamefont{Y.-A.} \bibnamefont{Chen}}, \bibnamefont{and}
  \bibinfo{author}{\bibfnamefont{I.}~\bibnamefont{Bloch}},
  \bibinfo{journal}{Phys. Rev. Lett.} \textbf{\bibinfo{volume}{107}},
  \bibinfo{pages}{255301} (\bibinfo{year}{2011}).

\bibitem[{\citenamefont{Struck et~al.}(2012)\citenamefont{Struck,
  \"Olschl\"ager, Weinberg, Hauke, Simonet, Eckardt, Lewenstein, Sengstock, and
  Windpassinger}}]{Struck2012}
\bibinfo{author}{\bibfnamefont{J.}~\bibnamefont{Struck}},
  \bibinfo{author}{\bibfnamefont{C.}~\bibnamefont{\"Olschl\"ager}},
  \bibinfo{author}{\bibfnamefont{M.}~\bibnamefont{Weinberg}},
  \bibinfo{author}{\bibfnamefont{P.}~\bibnamefont{Hauke}},
  \bibinfo{author}{\bibfnamefont{J.}~\bibnamefont{Simonet}},
  \bibinfo{author}{\bibfnamefont{A.}~\bibnamefont{Eckardt}},
  \bibinfo{author}{\bibfnamefont{M.}~\bibnamefont{Lewenstein}},
  \bibinfo{author}{\bibfnamefont{K.}~\bibnamefont{Sengstock}},
  \bibnamefont{and}
  \bibinfo{author}{\bibfnamefont{P.}~\bibnamefont{Windpassinger}},
  \bibinfo{journal}{Phys. Rev. Lett.} \textbf{\bibinfo{volume}{108}},
  \bibinfo{pages}{225304} (\bibinfo{year}{2012}).

\bibitem[{\citenamefont{Miyake et~al.}(2013)\citenamefont{Miyake, Siviloglou,
  Kennedy, Burton, and Ketterle}}]{Miyake2013}
\bibinfo{author}{\bibfnamefont{H.}~\bibnamefont{Miyake}},
  \bibinfo{author}{\bibfnamefont{G.~A.} \bibnamefont{Siviloglou}},
  \bibinfo{author}{\bibfnamefont{C.~J.} \bibnamefont{Kennedy}},
  \bibinfo{author}{\bibfnamefont{W.~C.} \bibnamefont{Burton}},
  \bibnamefont{and} \bibinfo{author}{\bibfnamefont{W.}~\bibnamefont{Ketterle}},
  \bibinfo{journal}{Phys. Rev. Lett.} \textbf{\bibinfo{volume}{111}},
  \bibinfo{pages}{185302} (\bibinfo{year}{2013}).

\bibitem[{\citenamefont{Aidelsburger et~al.}(2013)\citenamefont{Aidelsburger,
  Atala, Lohse, Barreiro, Paredes, and Bloch}}]{Aidelsburger2013}
\bibinfo{author}{\bibfnamefont{M.}~\bibnamefont{Aidelsburger}},
  \bibinfo{author}{\bibfnamefont{M.}~\bibnamefont{Atala}},
  \bibinfo{author}{\bibfnamefont{M.}~\bibnamefont{Lohse}},
  \bibinfo{author}{\bibfnamefont{J.~T.} \bibnamefont{Barreiro}},
  \bibinfo{author}{\bibfnamefont{B.}~\bibnamefont{Paredes}}, \bibnamefont{and}
  \bibinfo{author}{\bibfnamefont{I.}~\bibnamefont{Bloch}},
  \bibinfo{journal}{Phys. Rev. Lett.} \textbf{\bibinfo{volume}{111}},
  \bibinfo{pages}{185301} (\bibinfo{year}{2013}).

\bibitem[{\citenamefont{Oka and Aoki}(2009)}]{oka2009}
\bibinfo{author}{\bibfnamefont{T.}~\bibnamefont{Oka}} \bibnamefont{and}
  \bibinfo{author}{\bibfnamefont{H.}~\bibnamefont{Aoki}},
  \bibinfo{journal}{Phys. Rev. B} \textbf{\bibinfo{volume}{79}},
  \bibinfo{pages}{081406} (\bibinfo{year}{2009}).

\bibitem[{\citenamefont{Wang et~al.}(2013)\citenamefont{Wang, Steinberg,
  Jarillo-Herrero, and Gedik}}]{wang2013}
\bibinfo{author}{\bibfnamefont{Y.}~\bibnamefont{Wang}},
  \bibinfo{author}{\bibfnamefont{H.}~\bibnamefont{Steinberg}},
  \bibinfo{author}{\bibfnamefont{P.}~\bibnamefont{Jarillo-Herrero}},
  \bibnamefont{and} \bibinfo{author}{\bibfnamefont{N.}~\bibnamefont{Gedik}},
  \bibinfo{journal}{Science} \textbf{\bibinfo{volume}{342}},
  \bibinfo{pages}{453} (\bibinfo{year}{2013}).

\bibitem[{\citenamefont{Hafezi et~al.}(2011)\citenamefont{Hafezi, Demler,
  Lukin, and Taylor}}]{Hafezi2011}
\bibinfo{author}{\bibfnamefont{M.}~\bibnamefont{Hafezi}},
  \bibinfo{author}{\bibfnamefont{E.~A.} \bibnamefont{Demler}},
  \bibinfo{author}{\bibfnamefont{M.~D.} \bibnamefont{Lukin}}, \bibnamefont{and}
  \bibinfo{author}{\bibfnamefont{J.~M.} \bibnamefont{Taylor}},
  \bibinfo{journal}{Nat Phys} \textbf{\bibinfo{volume}{7}},
  \bibinfo{pages}{907} (\bibinfo{year}{2011}), ISSN \bibinfo{issn}{1745-2473}.

\bibitem[{\citenamefont{Rechtsman et~al.}(2013)\citenamefont{Rechtsman, Zeuner,
  Plotnik, Lumer, Podolsky, Dreisow, Nolte, Segev, and
  Szameit}}]{Rechtsman2013}
\bibinfo{author}{\bibfnamefont{M.~C.} \bibnamefont{Rechtsman}},
  \bibinfo{author}{\bibfnamefont{J.~M.} \bibnamefont{Zeuner}},
  \bibinfo{author}{\bibfnamefont{Y.}~\bibnamefont{Plotnik}},
  \bibinfo{author}{\bibfnamefont{Y.}~\bibnamefont{Lumer}},
  \bibinfo{author}{\bibfnamefont{D.}~\bibnamefont{Podolsky}},
  \bibinfo{author}{\bibfnamefont{F.}~\bibnamefont{Dreisow}},
  \bibinfo{author}{\bibfnamefont{S.}~\bibnamefont{Nolte}},
  \bibinfo{author}{\bibfnamefont{M.}~\bibnamefont{Segev}}, \bibnamefont{and}
  \bibinfo{author}{\bibfnamefont{A.}~\bibnamefont{Szameit}},
  \bibinfo{journal}{Nature} \textbf{\bibinfo{volume}{496}},
  \bibinfo{pages}{196} (\bibinfo{year}{2013}), ISSN \bibinfo{issn}{0028-0836}.

\bibitem[{\citenamefont{Mittal et~al.}(2016)\citenamefont{Mittal, Ganeshan,
  Fan, Vaezi, and Hafezi}}]{Mittal2016}
\bibinfo{author}{\bibfnamefont{S.}~\bibnamefont{Mittal}},
  \bibinfo{author}{\bibfnamefont{S.}~\bibnamefont{Ganeshan}},
  \bibinfo{author}{\bibfnamefont{J.}~\bibnamefont{Fan}},
  \bibinfo{author}{\bibfnamefont{A.}~\bibnamefont{Vaezi}}, \bibnamefont{and}
  \bibinfo{author}{\bibfnamefont{M.}~\bibnamefont{Hafezi}},
  \bibinfo{journal}{Nat Photon} \textbf{\bibinfo{volume}{10}},
  \bibinfo{pages}{180} (\bibinfo{year}{2016}), ISSN \bibinfo{issn}{1749-4885}.

\bibitem[{\citenamefont{Greschner et~al.}(2013)\citenamefont{Greschner, Santos,
  and Vekua}}]{Greschner2013}
\bibinfo{author}{\bibfnamefont{S.}~\bibnamefont{Greschner}},
  \bibinfo{author}{\bibfnamefont{L.}~\bibnamefont{Santos}}, \bibnamefont{and}
  \bibinfo{author}{\bibfnamefont{T.}~\bibnamefont{Vekua}},
  \bibinfo{journal}{Phys. Rev. A} \textbf{\bibinfo{volume}{87}},
  \bibinfo{pages}{033609} (\bibinfo{year}{2013}).

\bibitem[{\citenamefont{Mishra et~al.}(2014)\citenamefont{Mishra, Pai, and
  Mukerjee}}]{Mishra2014}
\bibinfo{author}{\bibfnamefont{T.}~\bibnamefont{Mishra}},
  \bibinfo{author}{\bibfnamefont{R.~V.} \bibnamefont{Pai}}, \bibnamefont{and}
  \bibinfo{author}{\bibfnamefont{S.}~\bibnamefont{Mukerjee}},
  \bibinfo{journal}{Phys. Rev. A} \textbf{\bibinfo{volume}{89}},
  \bibinfo{pages}{013615} (\bibinfo{year}{2014}).

\bibitem[{\citenamefont{Mishra et~al.}(2015{\natexlab{a}})\citenamefont{Mishra,
  Greschner, and Santos}}]{Mishra2015molecule}
\bibinfo{author}{\bibfnamefont{T.}~\bibnamefont{Mishra}},
  \bibinfo{author}{\bibfnamefont{S.}~\bibnamefont{Greschner}},
  \bibnamefont{and} \bibinfo{author}{\bibfnamefont{L.}~\bibnamefont{Santos}},
  \bibinfo{journal}{Phys. Rev. A} \textbf{\bibinfo{volume}{91}},
  \bibinfo{pages}{043614} (\bibinfo{year}{2015}{\natexlab{a}}).

\bibitem[{\citenamefont{Anisimovas et~al.}(2016)\citenamefont{Anisimovas,
  Ra\ifmmode \check{c}\else \v{c}\fi{}i\ifmmode~\bar{u}\else \={u}\fi{}nas,
  Str\"ater, Eckardt, Spielman, and Juzeli\ifmmode~\bar{u}\else
  \={u}\fi{}nas}}]{Anisimovas2016}
\bibinfo{author}{\bibfnamefont{E.}~\bibnamefont{Anisimovas}},
  \bibinfo{author}{\bibfnamefont{M.}~\bibnamefont{Ra\ifmmode \check{c}\else
  \v{c}\fi{}i\ifmmode~\bar{u}\else \={u}\fi{}nas}},
  \bibinfo{author}{\bibfnamefont{C.}~\bibnamefont{Str\"ater}},
  \bibinfo{author}{\bibfnamefont{A.}~\bibnamefont{Eckardt}},
  \bibinfo{author}{\bibfnamefont{I.~B.} \bibnamefont{Spielman}},
  \bibnamefont{and}
  \bibinfo{author}{\bibfnamefont{G.}~\bibnamefont{Juzeli\ifmmode~\bar{u}\else
  \={u}\fi{}nas}}, \bibinfo{journal}{Phys. Rev. A}
  \textbf{\bibinfo{volume}{94}}, \bibinfo{pages}{063632}
  (\bibinfo{year}{2016}).

\bibitem[{\citenamefont{Huber and Altman}(2010)}]{Huber2010}
\bibinfo{author}{\bibfnamefont{S.~D.} \bibnamefont{Huber}} \bibnamefont{and}
  \bibinfo{author}{\bibfnamefont{E.}~\bibnamefont{Altman}},
  \bibinfo{journal}{Phys. Rev. B} \textbf{\bibinfo{volume}{82}},
  \bibinfo{pages}{184502} (\bibinfo{year}{2010}).

\bibitem[{\citenamefont{Mishra et~al.}(2015{\natexlab{b}})\citenamefont{Mishra,
  Greschner, and Santos}}]{Mishra2015sawtooth}
\bibinfo{author}{\bibfnamefont{T.}~\bibnamefont{Mishra}},
  \bibinfo{author}{\bibfnamefont{S.}~\bibnamefont{Greschner}},
  \bibnamefont{and} \bibinfo{author}{\bibfnamefont{L.}~\bibnamefont{Santos}},
  \bibinfo{journal}{Phys. Rev. B} \textbf{\bibinfo{volume}{92}},
  \bibinfo{pages}{195149} (\bibinfo{year}{2015}{\natexlab{b}}).

\bibitem[{\citenamefont{Tonegawa and Harada}(1987)}]{Tonegawa1987}
\bibinfo{author}{\bibfnamefont{T.}~\bibnamefont{Tonegawa}} \bibnamefont{and}
  \bibinfo{author}{\bibfnamefont{I.}~\bibnamefont{Harada}},
  \bibinfo{journal}{J. Phys. Soc. Jpn.} \textbf{\bibinfo{volume}{56}},
  \bibinfo{pages}{2153} (\bibinfo{year}{1987}).

\bibitem[{\citenamefont{Sarkar and Sen}(2002)}]{Sarkarrrt2002}
\bibinfo{author}{\bibfnamefont{S.}~\bibnamefont{Sarkar}} \bibnamefont{and}
  \bibinfo{author}{\bibfnamefont{D.}~\bibnamefont{Sen}},
  \bibinfo{journal}{Phys. Rev. B} \textbf{\bibinfo{volume}{65}},
  \bibinfo{pages}{172408} (\bibinfo{year}{2002}).

\bibitem[{\citenamefont{Capriotti et~al.}(2003)\citenamefont{Capriotti, Becca,
  Sorella, and Parola}}]{Capriotti2003}
\bibinfo{author}{\bibfnamefont{L.}~\bibnamefont{Capriotti}},
  \bibinfo{author}{\bibfnamefont{F.}~\bibnamefont{Becca}},
  \bibinfo{author}{\bibfnamefont{S.}~\bibnamefont{Sorella}}, \bibnamefont{and}
  \bibinfo{author}{\bibfnamefont{A.}~\bibnamefont{Parola}},
  \bibinfo{journal}{Phys. Rev. B} \textbf{\bibinfo{volume}{67}},
  \bibinfo{pages}{172404} (\bibinfo{year}{2003}).

\bibitem[{\citenamefont{Nakane et~al.}(2006)\citenamefont{Nakane, Fukumoto, and
  Oguchi}}]{Nakane2006}
\bibinfo{author}{\bibfnamefont{M.}~\bibnamefont{Nakane}},
  \bibinfo{author}{\bibfnamefont{Y.}~\bibnamefont{Fukumoto}}, \bibnamefont{and}
  \bibinfo{author}{\bibfnamefont{A.}~\bibnamefont{Oguchi}},
  \bibinfo{journal}{J. Phys. Soc. Jpn.} \textbf{\bibinfo{volume}{75}},
  \bibinfo{pages}{114712} (\bibinfo{year}{2006}).

\bibitem[{\citenamefont{Sen et~al.}(1996)\citenamefont{Sen, Shastry, Walstedt,
  and Cava}}]{Sensawtooth1996}
\bibinfo{author}{\bibfnamefont{D.}~\bibnamefont{Sen}},
  \bibinfo{author}{\bibfnamefont{B.~S.} \bibnamefont{Shastry}},
  \bibinfo{author}{\bibfnamefont{R.~E.} \bibnamefont{Walstedt}},
  \bibnamefont{and} \bibinfo{author}{\bibfnamefont{R.}~\bibnamefont{Cava}},
  \bibinfo{journal}{Phys. Rev. B} \textbf{\bibinfo{volume}{53}},
  \bibinfo{pages}{6401} (\bibinfo{year}{1996}).

\bibitem[{\citenamefont{White}(1992)}]{White1992}
\bibinfo{author}{\bibfnamefont{S.~R.} \bibnamefont{White}},
  \bibinfo{journal}{Phys. Rev. Lett.} \textbf{\bibinfo{volume}{69}},
  \bibinfo{pages}{2863} (\bibinfo{year}{1992}).

\bibitem[{\citenamefont{Schollw{\"o}ck}(2011)}]{Schollwoeck2011}
\bibinfo{author}{\bibfnamefont{U.}~\bibnamefont{Schollw{\"o}ck}},
  \bibinfo{journal}{Annals of Physics} \textbf{\bibinfo{volume}{326}},
  \bibinfo{pages}{96} (\bibinfo{year}{2011}).

\bibitem[{\citenamefont{Calabrese and J.~Cardy}(2004)}]{Calabrese2004}
\bibinfo{author}{\bibfnamefont{P.}~\bibnamefont{Calabrese}} \bibnamefont{and}
  \bibinfo{author}{\bibfnamefont{J.}~\bibnamefont{J.~Cardy}},
  \bibinfo{journal}{J. Stat. Mech.: Theory Exp.} p. \bibinfo{pages}{P06002}
  (\bibinfo{year}{2004}).

\bibitem[{\citenamefont{Nersesyan et~al.}(1998)\citenamefont{Nersesyan,
  Gogolin, and E{\ss}ler}}]{Nersesyan1998}
\bibinfo{author}{\bibfnamefont{A.~A.} \bibnamefont{Nersesyan}},
  \bibinfo{author}{\bibfnamefont{A.~O.} \bibnamefont{Gogolin}},
  \bibnamefont{and} \bibinfo{author}{\bibfnamefont{F.~H.}
  \bibnamefont{E{\ss}ler}}, \bibinfo{journal}{Phys. Rev. Lett.}
  \textbf{\bibinfo{volume}{81}}, \bibinfo{pages}{910} (\bibinfo{year}{1998}).

\bibitem[{\citenamefont{Osterloh et~al.}(2002)\citenamefont{Osterloh, Amico,
  Falci, and Fazio}}]{osterloh2002}
\bibinfo{author}{\bibfnamefont{A.}~\bibnamefont{Osterloh}},
  \bibinfo{author}{\bibfnamefont{L.}~\bibnamefont{Amico}},
  \bibinfo{author}{\bibfnamefont{G.}~\bibnamefont{Falci}}, \bibnamefont{and}
  \bibinfo{author}{\bibfnamefont{R.}~\bibnamefont{Fazio}},
  \bibinfo{journal}{Nature} \textbf{\bibinfo{volume}{416}},
  \bibinfo{pages}{608} (\bibinfo{year}{2002}).

\bibitem[{\citenamefont{Vidal et~al.}(2003)\citenamefont{Vidal, Latorre, Rico,
  and Kitaev}}]{Vidal2003}
\bibinfo{author}{\bibfnamefont{G.}~\bibnamefont{Vidal}},
  \bibinfo{author}{\bibfnamefont{J.~I.} \bibnamefont{Latorre}},
  \bibinfo{author}{\bibfnamefont{E.}~\bibnamefont{Rico}}, \bibnamefont{and}
  \bibinfo{author}{\bibfnamefont{A.}~\bibnamefont{Kitaev}},
  \bibinfo{journal}{Phys. Rev. Lett.} \textbf{\bibinfo{volume}{90}},
  \bibinfo{pages}{227902} (\bibinfo{year}{2003}).

\end{thebibliography}

\end{document}